\documentclass[conference]{IEEEtran}
\IEEEoverridecommandlockouts
\usepackage{cite}
\usepackage{amsmath,amssymb}
\usepackage{graphicx}
\usepackage{tabularx}
\usepackage{booktabs}
\usepackage{float}
\usepackage{flushend}
\makeatletter
\newcommand{\linebreakand}{%
  \end{@IEEEauthorhalign}
  \hfill\mbox{}\par
  \mbox{}\hfill
  \begin{@IEEEauthorhalign}
}
\makeatother

\begin{document}

\title{SentTrack: Sentiment-Driven Bottleneck Detection\\in GitHub Issue Repositories}

\author{\IEEEauthorblockN{Xinyu Hu}
\IEEEauthorblockA{\textit{University of Tennessee}\\
Knoxville, TN, USA\\
xhu48@vols.utk.edu}
\and
\IEEEauthorblockN{Ali Behbahani}
\IEEEauthorblockA{\textit{University of Tennessee}\\
Knoxville, TN, USA\\
abehbaha@vols.utk.edu}
\and
\IEEEauthorblockN{Daniel Moon}
\IEEEauthorblockA{\textit{University of Tennessee}\\
Knoxville, TN, USA\\
dmoon4@vols.utk.edu}
\linebreakand
\IEEEauthorblockN{Yaren Dogan}
\IEEEauthorblockA{\textit{University of Tennessee}\\
Knoxville, TN, USA\\
yarddoga@vols.utk.edu}
\and
\IEEEauthorblockN{Nasir U. Eisty}
\IEEEauthorblockA{\textit{University of Tennessee}\\
Knoxville, TN, USA\\
neisty@utk.com}
}

\maketitle

\begin{abstract}
Software engineering teams increasingly depend on GitHub issue threads to coordinate work, report bugs, and negotiate technical decisions, yet most repository health tools focus on code metrics and ignore the conversational dynamics that drive or stall development. This paper presents SentTrack, a dual-lens framework for detecting socio-technical bottlenecks from GitHub issue discussions. Applied to the AvaloniaUI open-source repository across approximately 9,000 issue threads, the framework addresses three questions: how to automate workflow-inefficiency detection from real-time conversational data, whether sentiment signals can surface risk earlier than traditional label-based methods, and how to isolate human narrative from machine-generated noise in mixed-media issue text.

SentTrack combines two complementary pipelines. A horizontal pipeline translates raw issue reports into clean summaries using a large language model, extracts mid-level concern phrases, and clusters them through UMAP and HDBSCAN, producing 613 semantic clusters from the first 3,608 issues processed. A vertical pipeline applies the ABCDE collaborative interaction framework to classify each comment and infer thread-level outcomes. Across the full corpus, 49\% of threads ended in stagnation and only 13\% reached resolution, with the resolution gap identified as the dominant bottleneck signal. A weighted scoring engine that combines negativity, stagnation, resolution gap, and thread length gives maintainers an interpretable prioritization tool for high-friction discussions before they stall development.
\end{abstract}


\section{Introduction}

Software development teams increasingly rely on GitHub Issues to coordinate work, report bugs, and negotiate technical decisions~\cite{b2,b3}. These threads constitute a continuous, real-time record of both technical problems and the communication patterns that shape how projects evolve. Yet most repository health tools focus narrowly on code metrics, treating issue discussions as secondary. Teams still depend on periodic retrospectives or manually curated labels to surface friction, an approach that is reactive by nature and slow to catch emerging problems.

This paper investigates whether that gap can be closed through automated analysis of GitHub issue conversations. We focus on three research questions:

\begin{itemize}
    \item \textbf{RQ1:} How can an automated framework identify workflow inefficiencies using real-time conversational data from GitHub repositories?
    \item \textbf{RQ2:} Can sentiment analysis surface risk signals earlier than traditional label-based repository management?
    \item \textbf{RQ3:} How can human narrative be effectively isolated from machine-generated noise in mixed-media issue threads to ensure topic extraction reflects actual developer intent?
\end{itemize}

We analyze issue discussions from the AvaloniaUI open-source repository through a dual-lens approach: a horizontal pipeline for global topic modeling and a vertical pipeline for per-thread interaction analysis. Public repositories such as AvaloniaUI involve a mix of core maintainers and external contributors whose discussions span technical debt, workflow friction, feature prioritization, and documentation gaps.

Two characteristics of GitHub issues make automated analysis particularly challenging. First, repository labels such as \textit{bug}, \textit{enhancement}, and \textit{documentation} are frequently inconsistent, overly specific, or missing context about why issues remain unresolved. Labels like \textit{UI-fix} and \textit{Backend-refactor} may both stem from a deeper systemic cause, yet no native mechanism aggregates them into a higher-level insight. Second, GitHub issues are mixed-media documents in which stack traces, log dumps, and code snippets dominate text volume but carry no user intent, causing language models to surface topics rooted in code artifacts rather than developer concerns.

SentTrack addresses both characteristics. The horizontal pipeline removes machine noise through intent-focused LLM summarization before extracting and clustering topics. The vertical pipeline models each issue as a chronological interaction sequence, classifying comments under the ABCDE framework to detect whether discussions converge toward resolution or accumulate friction. Together, the two lenses capture both \textit{what} developers discuss and \textit{how} those discussions progress, producing interpretable bottleneck scores that maintainers can act on directly.

\section{Related Work}

This study draws on five bodies of literature: sentiment and emotion mining in software engineering, issue mining and machine-noise removal, semantic topic modeling, thread-level conversation analysis, and LLM-assisted developer tooling.

\subsection{Sentiment and Emotion Mining in Software Engineering}

Pang and Lee~\cite{pang2008opinion} established the foundational vocabulary of sentiment analysis and opinion mining, treating them as closely related tasks over subjective text. Liu~\cite{liu2012sentiment} formalized Aspect-Based Sentiment Analysis (ABSA), representing an opinion as a quintuple
\begin{equation}
o=(e,a,s,h,t),
\label{eq:absa_quintuple}
\end{equation}
where $e$ is the target entity, $a$ is the aspect, $s$ is sentiment orientation, $h$ is the opinion holder, and $t$ is time. This formulation is well suited to GitHub because a contributor may express negative sentiment about one aspect, such as rendering or input handling, while remaining neutral toward the project overall.

Affective signals have been identified in commits, pull requests, issue trackers, and security discussions~\cite{jurado2015sentiment,guzman2014commit,pletea2014security,ortu2015bullies,ortu2016emotional}. However, general-purpose sentiment tools perform poorly on technical text because words that appear negative in ordinary language may describe neutral debugging facts. SE-specific tools such as SentiStrength-SE, Senti4SD, and SentiCR address this mismatch, though comparative studies show that tool choice can still alter empirical conclusions~\cite{islam2018sentistrengthse,calefato2018senti4sd,ahmed2017senticr,jongeling2017negative,novielli2021assessment,imran2022sentimenttools,coutinho2024looksgood}. Accordingly, SentTrack treats negativity as one term in a multi-factor risk model rather than as a direct bottleneck signal:
\begin{equation}
B_i=f(N_i,S_i,R_i,L_i),
\label{eq:risk_model}
\end{equation}
where $N_i$ is negativity, $S_i$ is stagnation, $R_i$ is resolution gap, and $L_i$ is normalized thread length.

\subsection{Issue Mining, Machine Noise, and Pipeline Error}

GitHub labels provide useful metadata but are often sparse, inconsistent, or too coarse for repository-health analysis~\cite{colavito2024impact}. A further complication is the mixed-media nature of issue reports:
\begin{equation}
\text{Issue Text} = \text{Human Narrative} + \text{Code and Logs.}
\label{eq:mixed_media}
\end{equation}
Non-natural-language artifacts degrade NLP pipelines when they are not removed or normalized~\cite{hirsch2022denoising}, and pipeline studies demonstrate that early-stage errors propagate into later components~\cite{finkel2006cascading,caselli2015error}. These findings motivate SentTrack's translation stage, which converts raw issue text into intent-focused summaries before topic extraction begins.

\subsection{Semantic Summarization, Topic Extraction, and Clustering}

The horizontal component transforms raw issues through three sequential stages:
\begin{equation}
X_i \xrightarrow{\;T\;} S_i \xrightarrow{\;E\;} K_i \xrightarrow{\;C\;} z_i,
\label{eq:horizontal_chain}
\end{equation}
where $X_i$ is raw issue text, $S_i$ is a clean summary, $K_i$ is a set of extracted concern phrases, and $z_i$ is the cluster assignment. Evaluation metrics such as ROUGE~\cite{lin2004rouge}, factual-consistency scoring~\cite{kryscinski2020factcc}, SEAHORSE~\cite{clark2023seahorse}, and GPTScore~\cite{fu2023gptscore} motivate assessing summaries beyond surface fluency. Keyphrase extraction work~\cite{augenstein2017scienceie} motivates transforming long technical documents into compact concepts before clustering.

Topic modeling has been applied to large developer corpora such as Stack Overflow~\cite{barua2014developers}. Embedding-based approaches such as BERTopic~\cite{grootendorst2022bertopic} combine semantic encoders with density-based clustering. SentTrack follows this direction but separates topic extraction from clustering so that over-fragmentation can be diagnosed independently at each stage. UMAP preserves local and global neighborhood structure in low dimensions~\cite{mcinnes2018umap}, HDBSCAN identifies density-based clusters without a predefined count~\cite{campello2013density}, and DBCV provides an intrinsic measure of cluster quality~\cite{moulavi2014dbcv}.

\subsection{Thread-Level Conversation Analysis and LLM Assistance}

Stavrianou et al.~\cite{stavrianou2009discussion} integrated social-network analysis with opinion mining to model web forum dynamics, distinguishing user-based from opinion-based graph structures. Their work supports treating a GitHub issue as a chronological sequence of replies rather than a single document.

Ravi et al.~\cite{ravi2025threading} introduce a collaborative interaction framework structured around the ABCDE taxonomy: Agree, Build, Chatting, Different Perspective, and Elicit. SentTrack adapts this taxonomy to GitHub issue threads, where labels $A$ and $B$ indicate convergence, $D$ and $E$ indicate unresolved concern or open clarification, and $C$ represents weakly task-moving text.

LLMs support a broad range of software-engineering tasks but remain vulnerable to hallucination and missing project context~\cite{fan2023llmse}. Retrieval-augmented generation grounds model outputs in retrieved artifacts~\cite{lewis2020rag}, and studies of developer assistance tools show that RAG-based responses can match human helpfulness, though verbosity remains a limitation~\cite{correia2024devmentorai,correia2025firefox}. SentTrack therefore uses LLMs only in bounded, well-defined roles: summarization, concern phrase extraction, and cluster labeling.

\subsection{Positioning of SentTrack}

Prior work typically studies sentiment, issue classification, summarization, topic modeling, or conversation structure as independent problems. SentTrack combines these streams by simultaneously estimating
\begin{equation}
\underbrace{\text{what is discussed}}_{\text{horizontal topics}}
\quad \text{and} \quad
\underbrace{\text{how discussions evolve}}_{\text{vertical thread dynamics.}}
\label{eq:dual_lens}
\end{equation}
The contribution is an interpretable end-to-end pipeline that links technical topics, thread outcomes, and bottleneck risk scores into a unified maintainer signal.

\section{Approach}

SentTrack is a dual-lens framework for detecting socio-technical bottlenecks in GitHub repositories. For issue $i$, the input is
\begin{equation}
I_i=(t_i,b_i,\mathcal{C}_i,m_i), \qquad
\mathcal{C}_i=(c_{i1},\ldots,c_{in_i}),
\label{eq:issue_input}
\end{equation}
where $t_i$ is the issue title, $b_i$ is the body, $\mathcal{C}_i$ is the ordered comment sequence, and $m_i$ contains metadata such as state and closure time. The horizontal lens maps $(t_i,b_i)$ to topic clusters; the vertical lens maps $\mathcal{C}_i$ to a thread outcome. A bottleneck scoring engine then fuses both lenses with sentiment-derived risk into a single interpretable score.

\subsection{Pipeline Overview}

The two sub-pipelines process each issue in parallel:
\begin{equation}
(t_i,b_i) \xrightarrow{T_\theta} S_i
\xrightarrow{E_\theta} \mathcal{K}_i
\xrightarrow{g,\,U,\,H} q_i,
\qquad
\mathcal{C}_i \xrightarrow{h} \mathbf{a}_i \rightarrow y_i,
\label{eq:pipeline}
\end{equation}
where $T_\theta$ performs intent-focused translation, $S_i$ is the resulting clean summary, $E_\theta$ extracts concern phrases $\mathcal{K}_i$, $g$ embeds phrases, $U$ applies UMAP, $H$ applies HDBSCAN, $q_i$ is the cluster assignment, $h$ assigns ABCDE labels, $\mathbf{a}_i$ is the interaction-label sequence, and $y_i$ is the inferred thread outcome. Table~\ref{tab:pipeline} summarizes each stage.

\begin{table}[t]
\centering
\caption{SentTrack pipeline stages and outputs.}
\label{tab:pipeline}
\begin{tabularx}{\columnwidth}{p{0.22\columnwidth}X p{0.23\columnwidth}}
\toprule
\textbf{Stage} & \textbf{Operation} & \textbf{Output} \\
\midrule
Translation & Remove machine noise while preserving user intent & $S_i$ \\
Topic extraction & Generate mid-level concern phrases using the project profile & $\mathcal{K}_i$ \\
Clustering & Embed, reduce, and density-cluster topics & $q_i$ \\
Vertical analysis & Classify comments as an ABCDE sequence & $\mathbf{a}_i, y_i$ \\
Scoring & Fuse negativity, stagnation, resolution gap, and thread length & $B_i$ \\
\bottomrule
\end{tabularx}
\end{table}

\subsection{Horizontal Topic Clustering}

The horizontal pipeline identifies macro-level repository trends and recurring technical themes across the issue corpus.

\subsubsection{Intent-Focused Summarization}

GitHub issues blend human narrative with machine-layer content such as stack traces and log dumps. SentTrack first produces a clean summary
\begin{equation}
S_i = T_\theta(t_i, b_i; P),
\label{eq:summarization}
\end{equation}
where $P$ is a structured project profile. The LLM is instructed to retain the developer's intent, normalize terminology, and suppress file paths, hexadecimal codes, stack traces, and raw logs. This prevents machine artifacts from dominating downstream topic extraction.

\subsubsection{Context-Aware Topic Extraction}

From each clean summary, the pipeline extracts a set of mid-level concern phrases:
\begin{equation}
\mathcal{K}_i = E_\theta(S_i; P) = \{k_{i1}, \ldots, k_{ir_i}\},
\label{eq:extraction}
\end{equation}
where each $k_{ij}$ is a three-to-six-word noun phrase. The project profile guides the model toward domain-specific concerns such as rendering, cross-platform input, data binding, tooling, documentation, performance, and native OS integration. The target abstraction level satisfies
\begin{equation}
\text{label} \;<\; k_{ij} \;<\; \text{raw stack trace,}
\label{eq:abstraction}
\end{equation}
meaning each phrase is more specific than a GitHub label but more reusable than a one-off error token.

\subsubsection{Embedding and Density-Based Clustering}

Each extracted phrase is embedded and assigned to a cluster through
\begin{equation}
\mathbf{v}_{ij} = g(k_{ij}), \qquad
\mathbf{u}_{ij} = U(\mathbf{v}_{ij}), \qquad
q_{ij} = H(\mathbf{u}_{ij}),
\label{eq:clustering}
\end{equation}
where $g$ is the instruction-tuned encoder, $U$ is UMAP, and $H$ is HDBSCAN. Because the number of recurring concerns in any repository is not known in advance, HDBSCAN is preferred over clustering methods that require a predefined count. The noise label $q_{ij} = -1$ is treated as informative: it marks rare or isolated concerns rather than a clustering failure.

\subsection{Vertical Thread Analysis}

The vertical pipeline treats each issue as a chronological discussion rather than a single block of text. For comment $c_{ij}$, a classifier assigns an ABCDE label:
\begin{equation}
a_{ij} = h(c_{ij}) \in \{A, B, C, D, E\}, \qquad
\mathbf{a}_i = (a_{i1}, \ldots, a_{in_i}).
\label{eq:abcde_assign}
\end{equation}
Table~\ref{tab:abcde} defines each label and its thread-level interpretation.

\begin{table}[t]
\centering
\caption{ABCDE coding framework for vertical thread analysis.}
\label{tab:abcde}
\begin{tabularx}{\columnwidth}{c p{0.27\columnwidth} X}
\toprule
\textbf{Code} & \textbf{Meaning} & \textbf{Thread signal} \\
\midrule
A & Agree & Confirmation or convergence \\
B & Build & Added evidence, proposed fix, or explanation \\
C & Chat & Neutral or weakly task-moving text \\
D & Different Perspective & Disagreement, concern, or blocker \\
E & Elicit & Question or clarification request \\
\bottomrule
\end{tabularx}
\end{table}

Thread-level counts and proportions are computed as
\begin{align}
n_i^{(r)} &= \sum_{j=1}^{n_i} \mathbf{1}[a_{ij} = r], \label{eq:abcde_count} \\
\rho_i^{(r)} &= \frac{n_i^{(r)}}{n_i}, \qquad r \in \{A, B, C, D, E\}. \label{eq:abcde_ratio}
\end{align}
The thread outcome is determined by the final interaction in the sequence:
\begin{equation}
y_i =
\begin{cases}
\text{resolved}, & a_{in_i} \in \{A, B\}, \\
\text{stagnant}, & a_{in_i} \in \{D, E\}, \\
\text{neutral},  & a_{in_i} = C.
\end{cases}
\label{eq:outcome}
\end{equation}
A sequence such as $E \rightarrow B \rightarrow A$ signals a question followed by a constructive response and agreement; $D \rightarrow E$ signals a concern followed by an unresolved clarification request.

\subsection{Pipeline Diagnostic Metrics}

To distinguish weak downstream NLP signals caused by model limitations from those caused by sparse or fragmented source conversations, SentTrack computes four diagnostic rates over $N$ threads:
\begin{align}
\mathrm{RG}  &= 1 - \frac{1}{N}\sum_{i=1}^{N} \mathbf{1}[y_i = \text{resolved}], \label{eq:rg} \\
\mathrm{QRR} &= \frac{\#(E \rightarrow \{A,B\})}{\#E}, \label{eq:qrr} \\
\mathrm{CCR} &= \frac{\#(\{D,E\} \rightarrow \{A,B\})}{\#\{D,E\}}, \label{eq:ccr} \\
\mathrm{FS}  &= \frac{\#\text{shallow or isolated threads}}{N}. \label{eq:fs}
\end{align}
Here RG is the resolution gap, QRR is the question-resolution rate, CCR is the concern-capture rate, and FS is the fragmentation score. High RG or low QRR and CCR indicates that issue threads provide weak conversational signals for downstream summarization and clustering, a property of the corpus rather than of the models.

\subsection{Bottleneck Detection Engine}

Rather than relying on a single indicator, the bottleneck engine combines sentiment-based features with thread-dynamics features. Comment-level negativity is aggregated per thread as
\begin{equation}
N_i = 0.50\,\overline{n}_i + 0.30\,n_i^{\max} + 0.20\,r_i^{-},
\label{eq:negativity}
\end{equation}
where $\overline{n}_i$ is average negativity, $n_i^{\max}$ is maximum negativity, and $r_i^{-}$ is the negative-comment ratio. The final bottleneck score is
\begin{equation}
B_i = 0.35 N_i + 0.30 S_i + 0.25 R_i + 0.10 L_i,
\label{eq:bottleneck}
\end{equation}
where $S_i$ is a binary stagnation indicator, $R_i$ is the resolution-gap score, and $L_i$ is normalized thread length. The weight on length is deliberately smallest because a long discussion may reflect legitimate technical complexity rather than communication failure.

In addition to the scalar score, the engine assigns each issue a compound bottleneck category:
\begin{equation}
\mathcal{G}_i \subseteq \{\text{Negative},\, \text{Stagnant},\, \text{Long},\, \text{Unresolved}\}.
\label{eq:categories}
\end{equation}
The score $B_i$ is intended as a prioritization signal for maintainer triage, not as a ground-truth label that an issue is objectively blocked.

\section{Implementation}

\subsection{Horizontal Analysis}

\subsubsection{Project Profiling}

To ensure the downstream LLM extracts domain-relevant topics rather than generic keywords, the pipeline begins with a static project profile stored in \texttt{project\_profile.md}. This profile was generated automatically by directing an AI agent to process the full AvaloniaUI repository and produce a structured document covering: the project's domain and typical users, its major user-facing features, the technology categories most likely to attract opinions, and the kinds of concerns users would plausibly raise in issue threads. The resulting profile defines AvaloniaUI as a cross-platform .NET UI framework and organizes expected feedback into technology pillars such as UI and Rendering, Performance and Resource Usage, and Tooling Integration, along with concrete representative concerns such as memory leaks, visual inconsistencies, and missing native OS integration.

\subsubsection{Intent-Focused Summarization}

Each raw issue title and body is processed by \texttt{gemini-2.5-flash} via \texttt{langchain\_google\_genai}, configured at temperature 0.2 to minimize hallucination. The model is prompted to act as an expert triage engineer: it strips machine-layer noise including file paths, memory addresses, hexadecimal codes, and stack traces, and produces a two-to-five sentence plain-English summary focused on the developer's core intent and the technical failure they experienced. A Pydantic schema enforces consistent output structure across the corpus.

\subsubsection{Topic Extraction}

Clean summaries are passed back through \texttt{gemini-2.5-flash} with a secondary Pydantic schema enforcing noun-phrase output of three to six words per phrase. This constraint keeps topics at the intended abstraction level, avoiding phrases that are too broad (such as ``Performance Bug'') or too specific (such as ``Slider track right side broken''). Extracted topics are deduplicated across the corpus to build a compact set of unique concern phrases for embedding.

\subsubsection{Semantic Embedding and Clustering}

Unique concern phrases are embedded with \texttt{Alibaba-NLP/gte-Qwen2-1.5B-instruct} via SentenceTransformers, producing high-dimensional vectors that capture contextual meaning rather than surface keyword overlap. UMAP reduces the embedding space while preserving local and global structure. HDBSCAN is then applied to the reduced representations, identifying natural-density clusters without a predefined count and labeling isolated phrases as noise.

\subsection{Vertical Thread Classification}

\subsubsection{Data Inputs}

The vertical pipeline processes scraped GitHub issue data containing issue identifiers, comment text, timestamps, issue state, and closure metadata where available. The full dataset covers approximately 9,000 parent issues, both open and closed. Comments are organized by issue number, sorted by creation timestamp, and analyzed at two levels: individually for interaction signals and collectively for thread-level aggregation.

\subsubsection{ABCDE Classification}

Each comment is converted to lowercase and matched against keyword lists for each interaction category. Phrases such as ``I agree,'' ``looks good,'' and ``that works'' are classified as A; proposed fixes, added evidence, and explanations as B; question marks and interrogative phrases such as ``why,'' ``could you,'' and ``what if'' as E; and expressions of concern or disagreement such as ``however,'' ``this breaks,'' and ``not sure'' as D. Comments not matching any category default to C. This rule-based approach provides a lightweight, transparent baseline for classifying comment roles without the computational cost of a large language model.

\subsubsection{Thread-Level Metrics}

After classification, the pipeline aggregates ABCDE labels into per-issue metrics: total comment count, per-label frequency, D and E ratios, a question-presence flag, the full chronological ABCDE sequence, and the thread outcome derived from Eq.~\ref{eq:outcome}. A transition matrix records how frequently each label type follows another across the corpus, enabling analysis of common conversation flow patterns.

\subsection{Bottleneck Detection}

The bottleneck engine merges thread-dynamics metrics from the vertical pipeline with comment-level sentiment scores produced by the VADER analyzer~\cite{hutto2014vader}. VADER assigns a negative score, a positive score, a neutral score, and a compound score to each comment independently. Per-thread aggregates including average negativity, maximum negativity, and negative-comment ratio are combined into $N_i$ via Eq.~\ref{eq:negativity}. Thread length is normalized by min-max scaling over all issues. Boolean flags for high negativity, stagnant outcome, long thread, and low resolution are derived and combined into the compound label set $\mathcal{G}_i$. The final bottleneck score $B_i$ from Eq.~\ref{eq:bottleneck} ranks all issues, with open and closed threads producing independent output files and visualization directories.

\section{Evaluation}

\subsection{Horizontal Topic Clustering}

Evaluating unsupervised NLP pipelines is inherently challenging without objective ground-truth labels. Because SentTrack intentionally replaces GitHub's fragile manual tagging system, no pre-existing baseline is available for direct comparison. We therefore evaluate the horizontal pipeline through qualitative assessment of individual extraction accuracy and quantitative analysis of discovery rates at scale.

\subsubsection{Qualitative Assessment}

On individual issues, human inspection confirms that the summarization and extraction constraints capture the core socio-technical friction. For a complex discussion about \texttt{INotifyPropertyChanged} and WPF \texttt{DataContext} equivalents, the pipeline generated clean mid-level phrases such as ``External Model Data Binding Implementation'' and ``MVVM Library Integration Guidance,'' successfully distilling a highly technical thread into structural concepts accessible to a project manager.

\subsubsection{Topic Fragmentation Analysis}

\begin{figure}[htbp]
    \centering
    \includegraphics[width=\columnwidth]{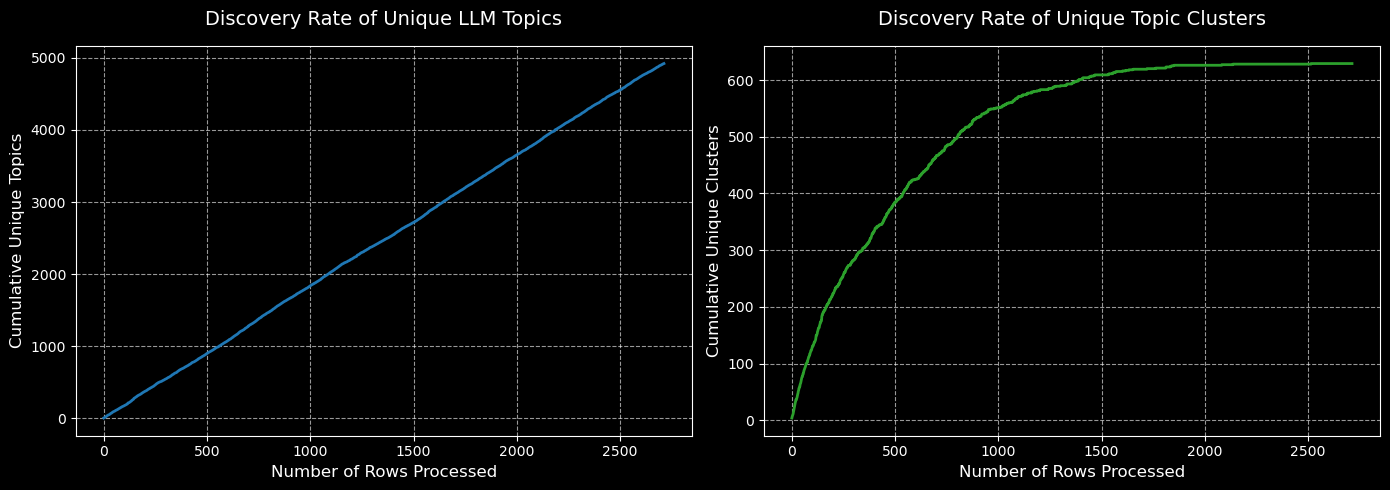}
    \caption{Discovery rates of unique LLM topics and HDBSCAN clusters across the first 2,700 issues.}
    \label{fig:discovery_partial}
\end{figure}

\begin{figure}[htbp]
    \centering
    \includegraphics[width=\columnwidth]{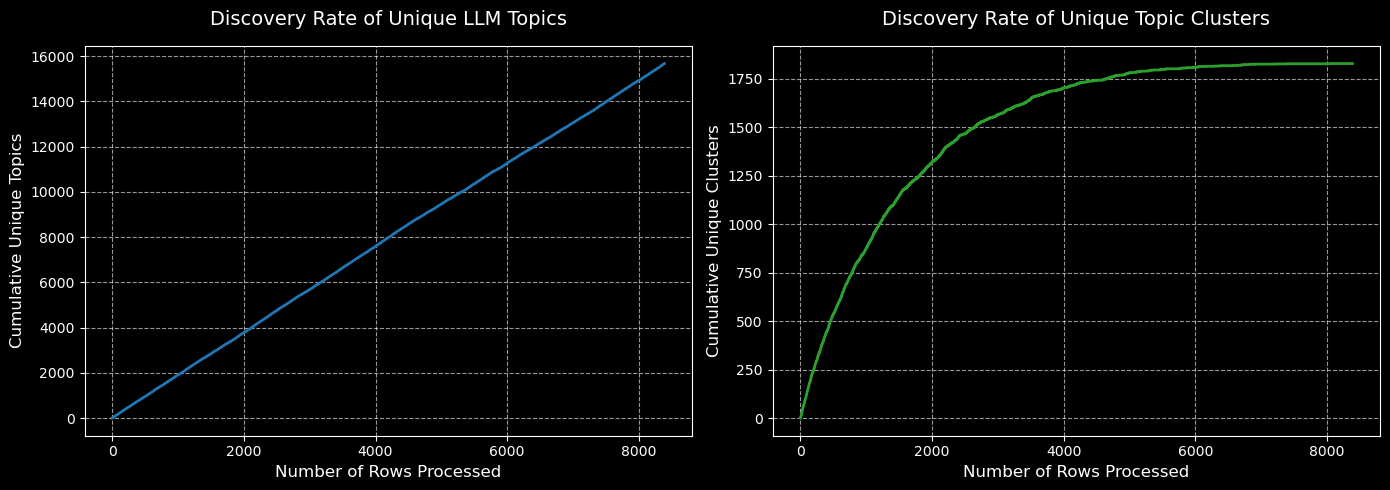}
    \caption{Discovery rates of unique LLM topics and HDBSCAN clusters across the full corpus of 6,629 issues.}
    \label{fig:discovery_full}
\end{figure}

To evaluate pipeline behavior at scale, we plotted the cumulative discovery rate of unique LLM-generated topics and unique HDBSCAN clusters against the number of issues processed. Figure~\ref{fig:discovery_partial} shows results for the first 2,700 issues and Figure~\ref{fig:discovery_full} shows results for the full corpus of 6,629 issues.

The unique-topic curve is nearly linear: processing approximately 8,000 rows yielded over 16,000 unique phrases, roughly two novel phrasings per issue. If the model produced consistent terminology, this curve would eventually plateau. Instead, the LLM generates fresh phrasings for essentially every issue, producing variants such as ``XAML previewer startup failure,'' ``designer process crash during initialization,'' and ``visual studio design view launch error'' for the same underlying problem.

The cluster curve behaves differently, exhibiting an S-shape that plateaus. This confirms that the semantic encoder and HDBSCAN do group many varied phrasings together. However, the plateau settles at approximately 620 clusters for 2,700 issues and over 1,800 clusters for the full corpus. A repository of AvaloniaUI's scale realistically contains 50 to 200 distinct macro-level concerns, so the pipeline produces roughly an order of magnitude more clusters than are actionable for maintainers.

\begin{figure}[htbp]
    \centering
    \includegraphics[width=\columnwidth]{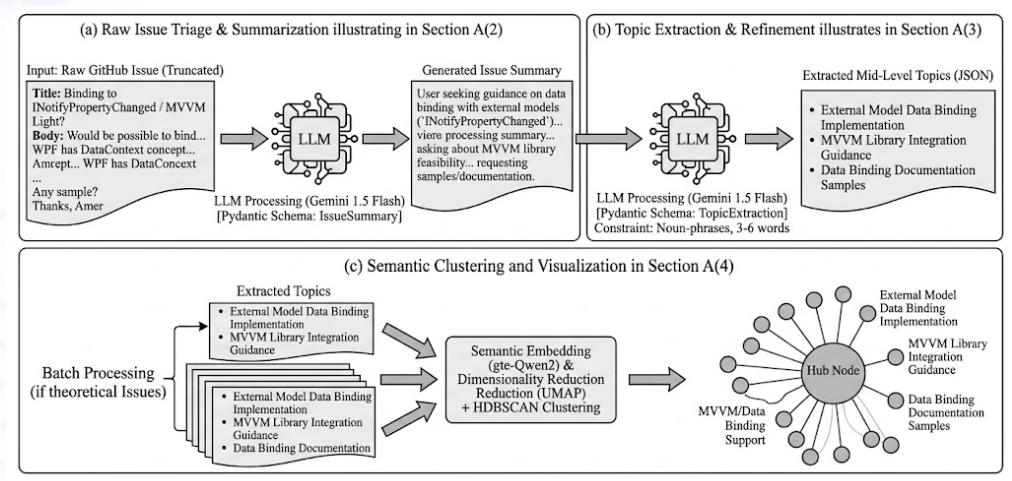}
    \caption{Representative horizontal pipeline execution showing: (a) LLM noise removal and summarization, (b) structured mid-level phrase extraction, and (c) mathematical cluster assignment.}
    \label{fig:pipeline_example}
\end{figure}

Figure~\ref{fig:pipeline_example} illustrates the full horizontal pipeline on a representative issue, showing how noise removal, phrase extraction, and cluster assignment interact.

The root cause of over-fragmentation is the LLM's unbounded extraction vocabulary. While the encoder can bridge moderate phrasing variation, linking ``dependency injection'' to ``IoC container'' for example, it cannot reliably consolidate 16,000 highly specific variants. Constraining the LLM to a curated vocabulary during extraction or introducing a dedicated LLM-driven topic-reduction step between clustering and reporting are the two most promising paths forward, and we treat this as the primary open problem in the horizontal pipeline.

\subsection{Vertical Thread Analysis}

\begin{figure}[H]
    \centering
    \includegraphics[width=0.75\linewidth]{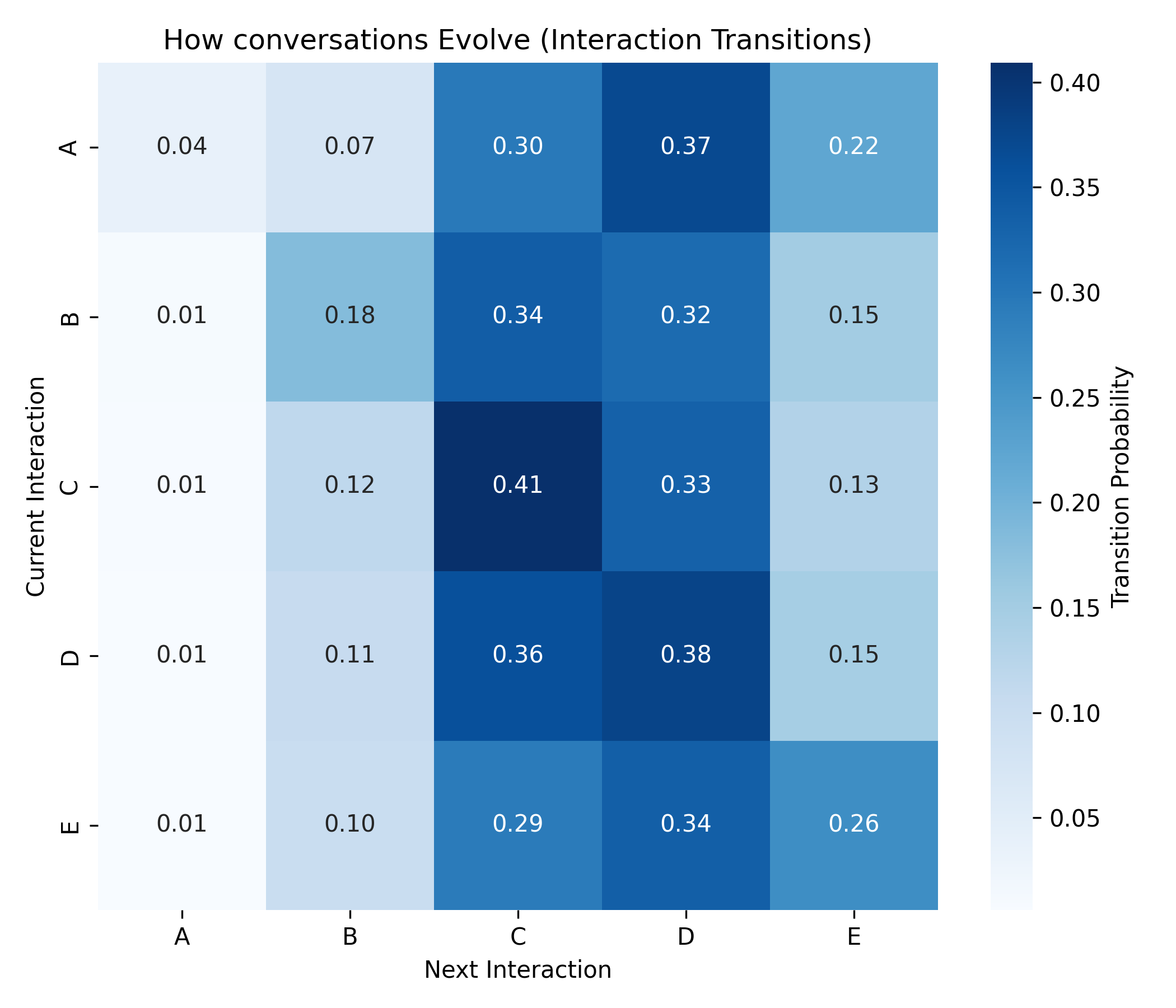}
    \caption{Interaction transition heatmap for closed issue threads.}
    \label{fig:heatmap_closed}
\end{figure}

\begin{figure}[H]
    \centering
    \includegraphics[width=0.75\linewidth]{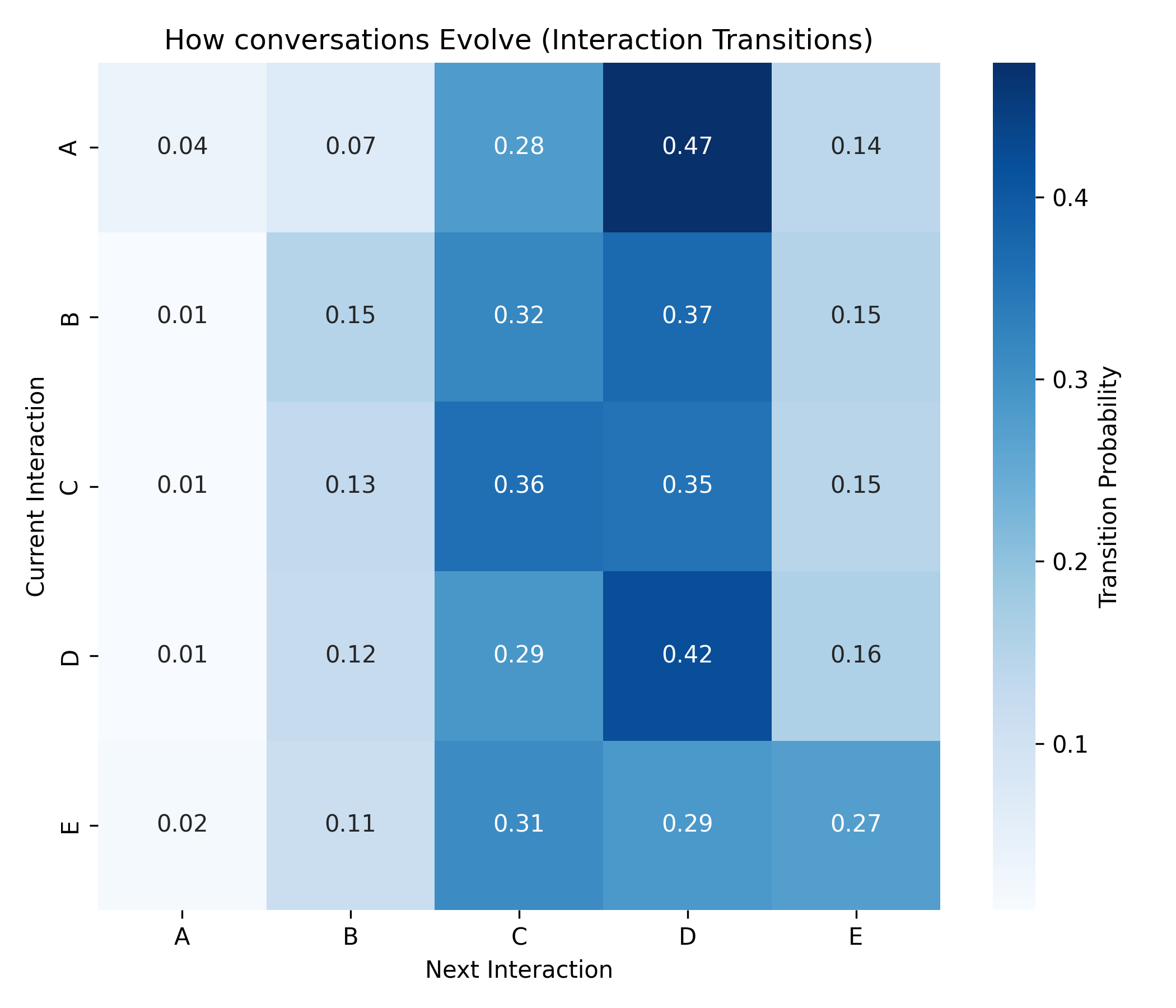}
    \caption{Interaction transition heatmap for open issue threads.}
    \label{fig:heatmap_open}
\end{figure}

\begin{figure}[H]
    \centering
    \includegraphics[width=0.75\linewidth]{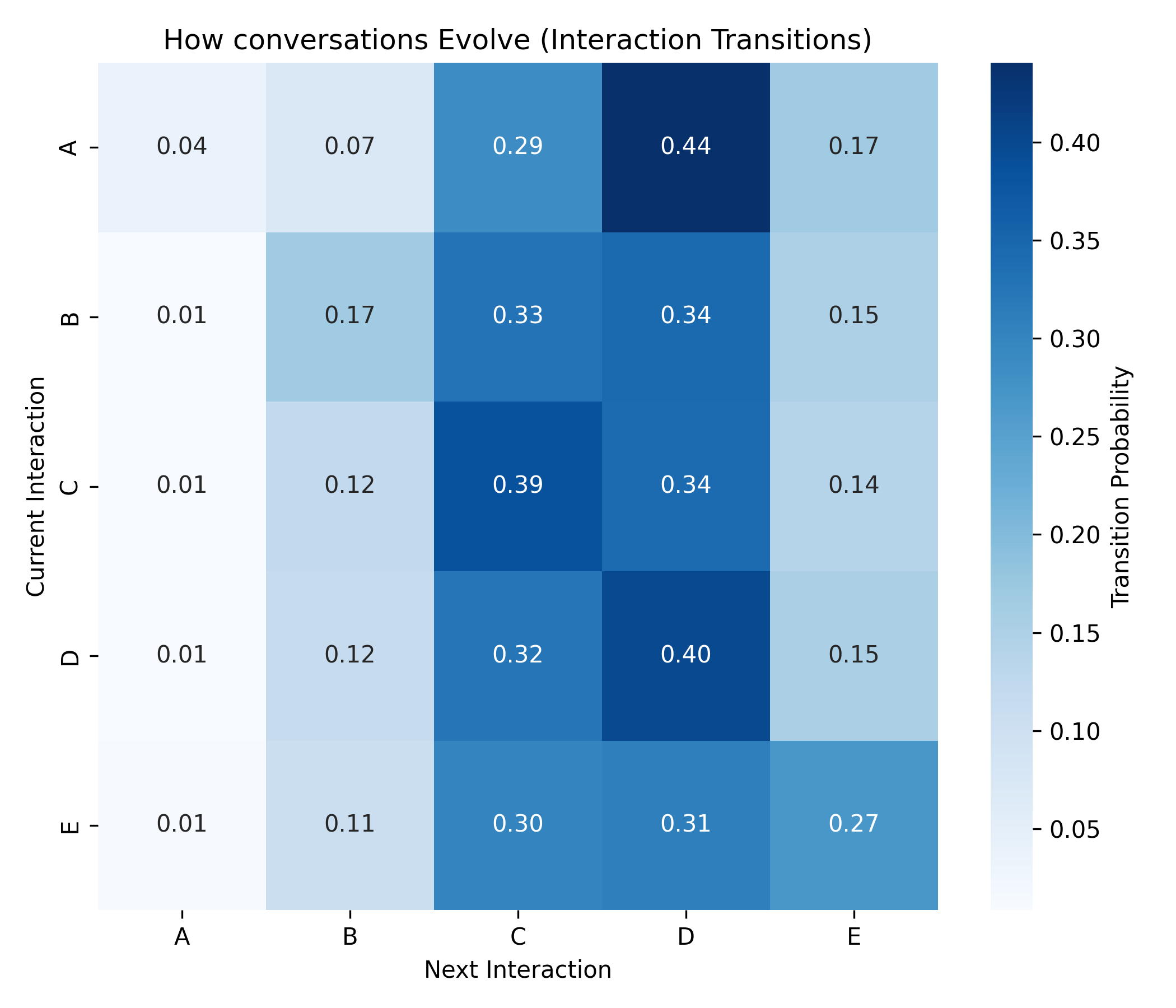}
    \caption{Interaction transition heatmap for all issue threads combined.}
    \label{fig:heatmap_combined}
\end{figure}

The transition heatmaps in Figures~\ref{fig:heatmap_closed}, \ref{fig:heatmap_open}, and \ref{fig:heatmap_combined} show interaction label transitions for closed threads, open threads, and the combined corpus respectively.

The ABCDE classifier reveals that most threads do not reach resolution: the majority end in disagreement, elicitation, or neutral commentary with no indication of subsequent progress. The resolution pattern differs only marginally between open and closed issues, so the combined corpus serves as the primary analytical unit.

\begin{figure}[H]
    \centering
    \includegraphics[width=0.75\linewidth]{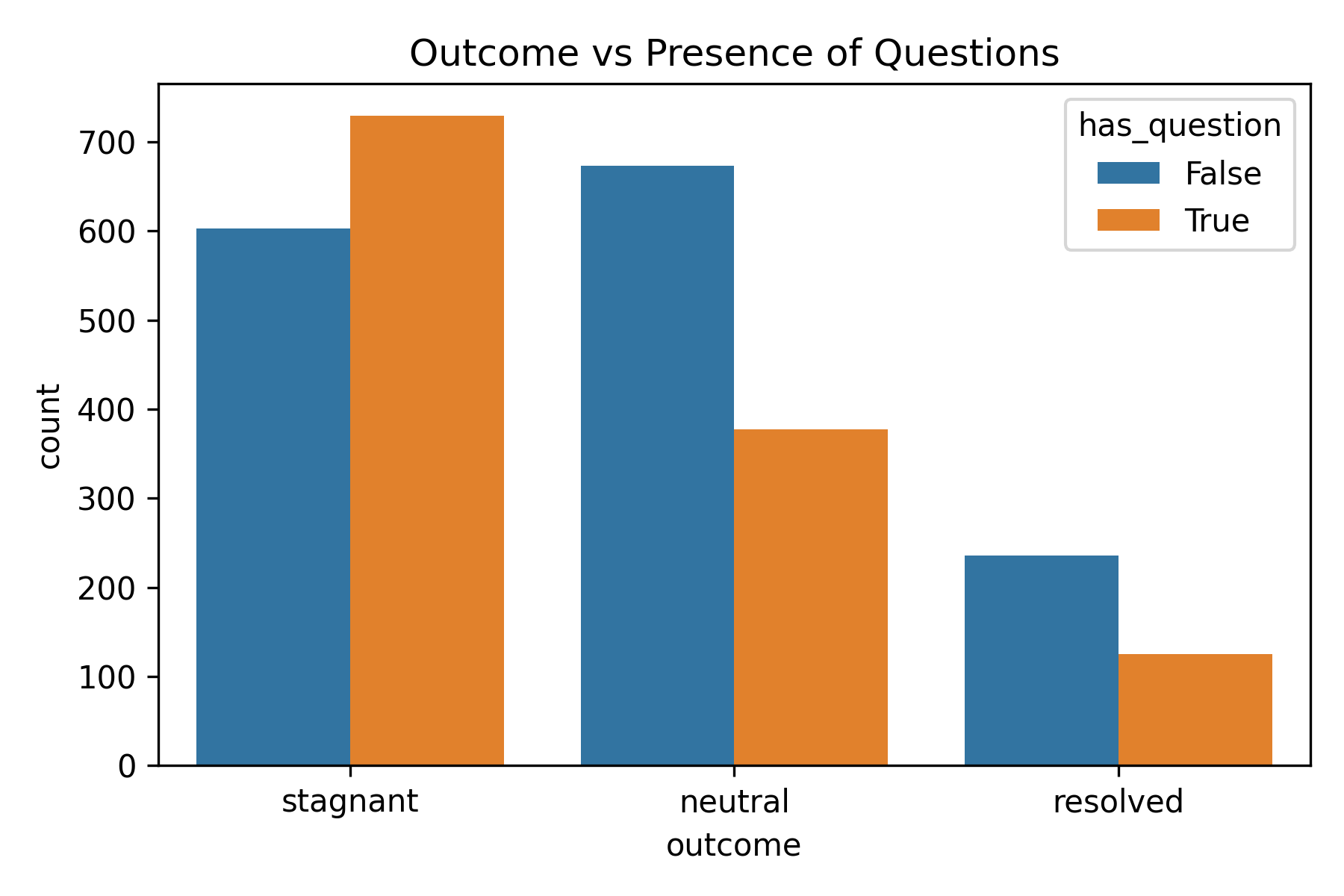}
    \caption{Thread outcome distribution conditioned on the presence of at least one elicitation comment.}
    \label{fig:outcome_questions}
\end{figure}

Across all threads, 49\% ended in stagnation (D or E), 38\% in neutral commentary (C), and 13\% in resolution (A or B). Figure~\ref{fig:outcome_questions} shows that the presence of questions is associated with stagnant outcomes: many resolved and neutral threads contain no elicitation at all.

\begin{figure}[H]
    \centering
    \includegraphics[width=0.75\linewidth]{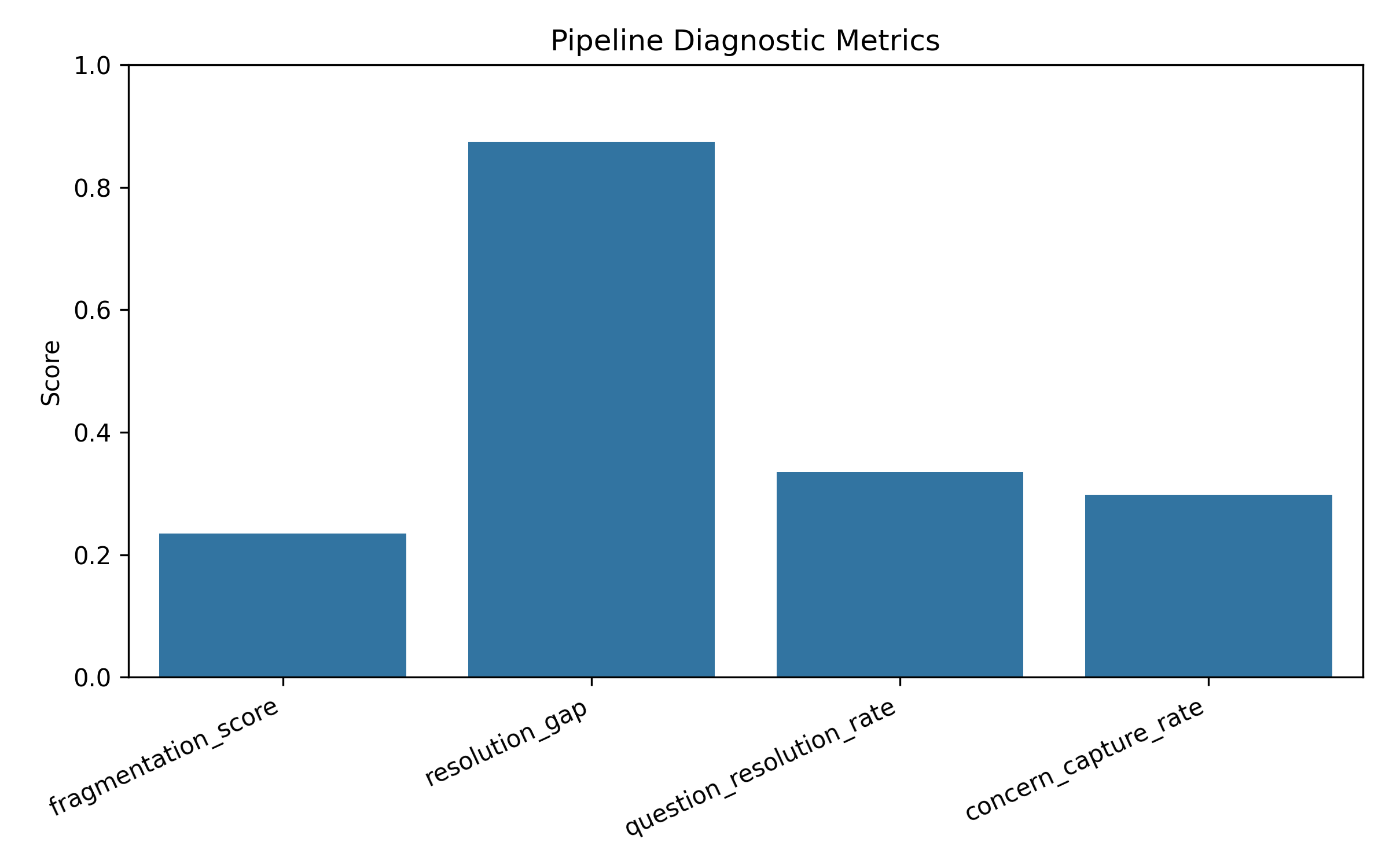}
    \caption{Pipeline diagnostic metrics over the combined open and closed corpus.}
    \label{fig:diagnostic}
\end{figure}

Figure~\ref{fig:diagnostic} shows the four diagnostic metrics computed over the full corpus. The resolution gap is the highest-scoring metric, confirming that unresolved discussions are the dominant bottleneck signal. The question-resolution rate and concern-capture rate are lower but meaningful, indicating that questions and concerns appear in threads but are not the primary friction source. The low fragmentation score suggests that conversations are not generally shallow or scattered; bottlenecks arise from resolution difficulty rather than from conversational disorganization.

\subsection{Bottleneck Detection Engine}

The bottleneck detection engine is evaluated through diagnostic visualizations that examine whether the pipeline identifies meaningful patterns of communication friction. Rather than treating detection as a supervised classification task, the evaluation focuses on interpretability and internal consistency between the scoring components.

\begin{figure}[H]
    \centering
    \includegraphics[width=1.00\linewidth]{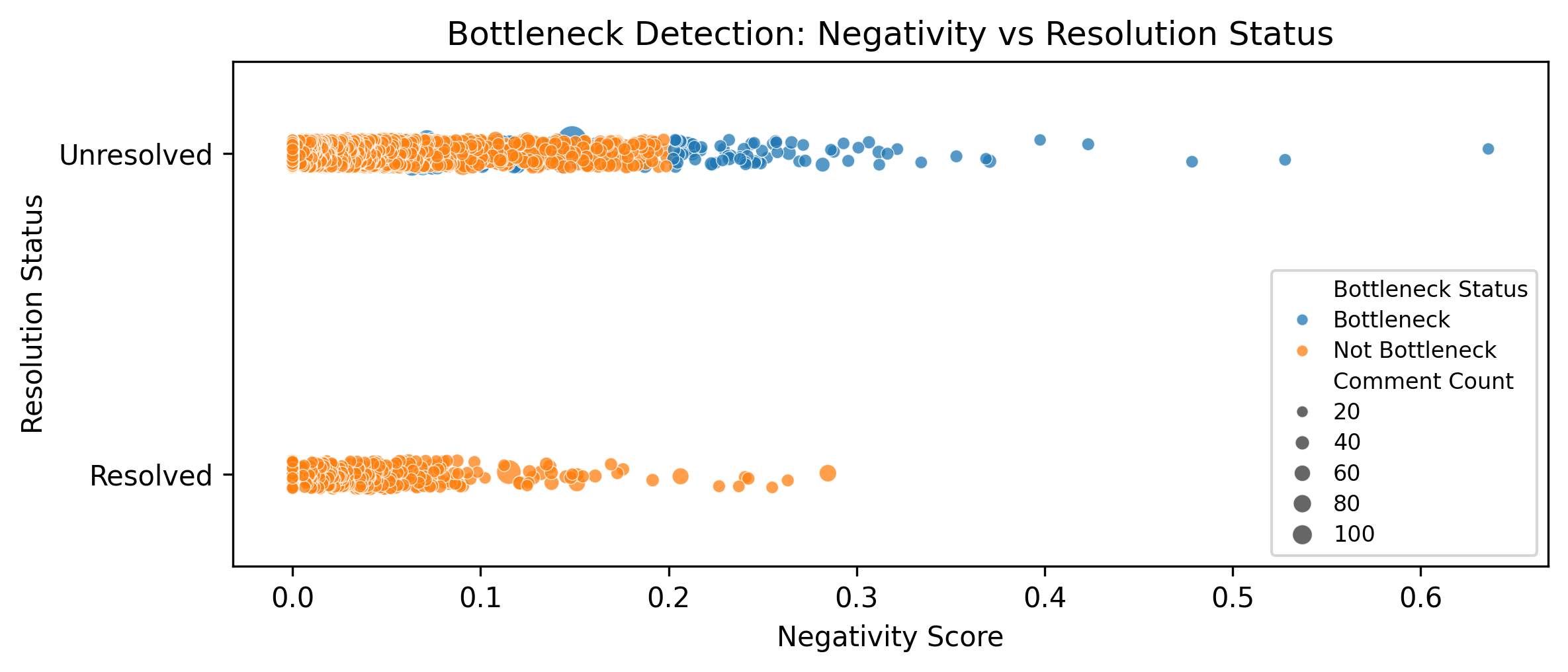}
    \caption{Issue negativity score versus resolution status, colored by bottleneck classification. Point size reflects comment count.}
    \label{fig:neg_resolution}
\end{figure}

Figure~\ref{fig:neg_resolution} compares each issue's negativity score against its resolution status, with points colored by bottleneck classification. Resolved issues cluster at low negativity and are predominantly classified as non-bottlenecks. Unresolved issues show a wider spread of negativity values and account for most detected bottlenecks, particularly as negativity exceeds the lower range. Point sizes reflect comment counts, and larger, more negative, unresolved threads tend to receive higher bottleneck scores. Negativity alone, however, does not characterize most unresolved threads, which underscores the importance of the multi-factor scoring approach.

\begin{figure}[H]
    \centering
    \includegraphics[width=1.00\linewidth]{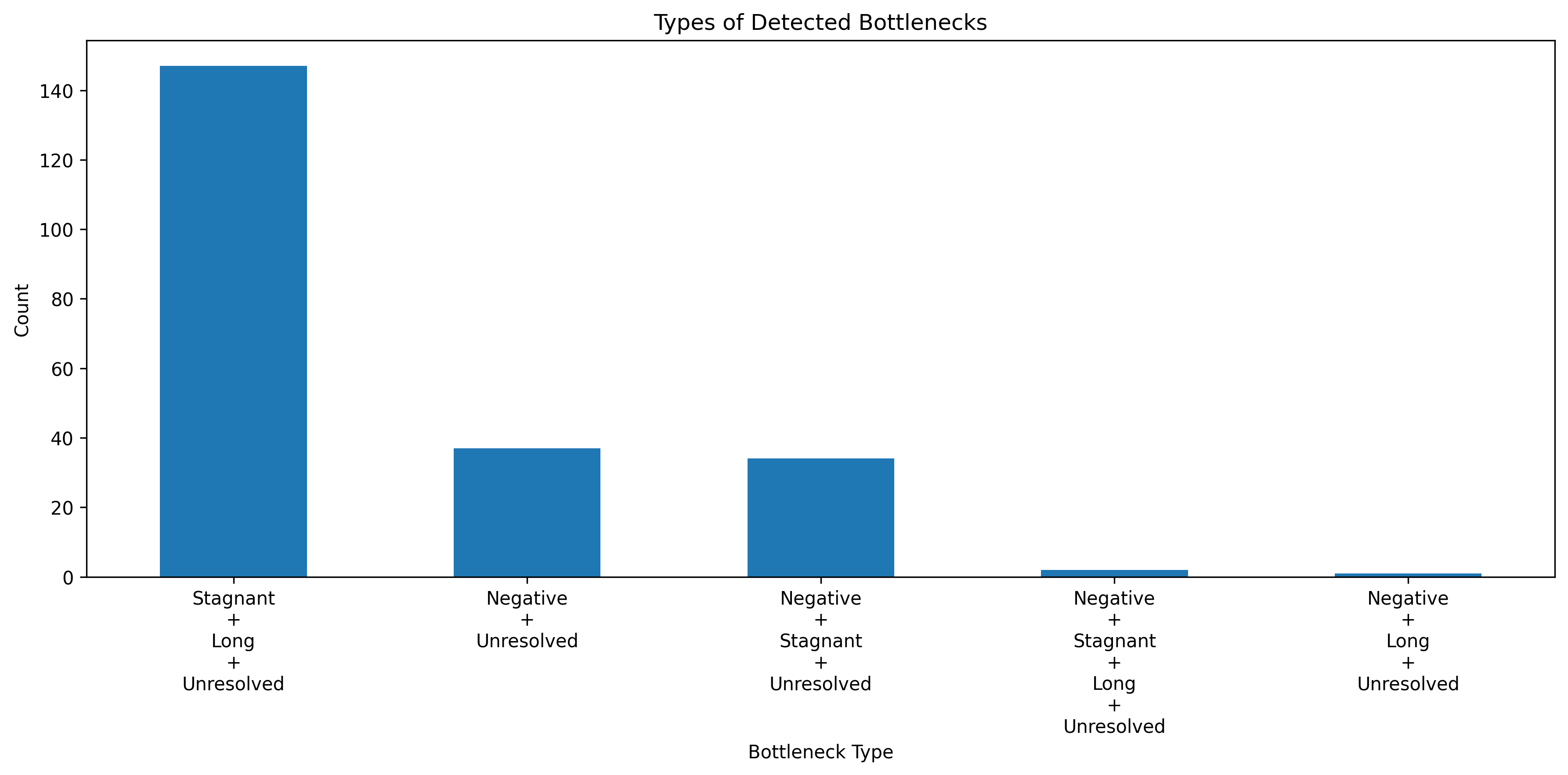}
    \caption{Distribution of compound bottleneck types across the issue corpus.}
    \label{fig:types}
\end{figure}

Figure~\ref{fig:types} shows the distribution of compound bottleneck categories across the corpus. The most common pattern is Stagnant and Long and Unresolved, confirming that the dominant signal is not emotional negativity but prolonged, directionless discussion. The categories Negative and Unresolved and Negative and Stagnant and Unresolved are the next most frequent, indicating that negative language contributes to bottleneck detection primarily when paired with unresolved outcomes. Multi-factor combinations involving all four signals are rare, suggesting that severe simultaneous failures are uncommon in the corpus.

\begin{figure}[H]
    \centering
    \includegraphics[width=1.00\linewidth]{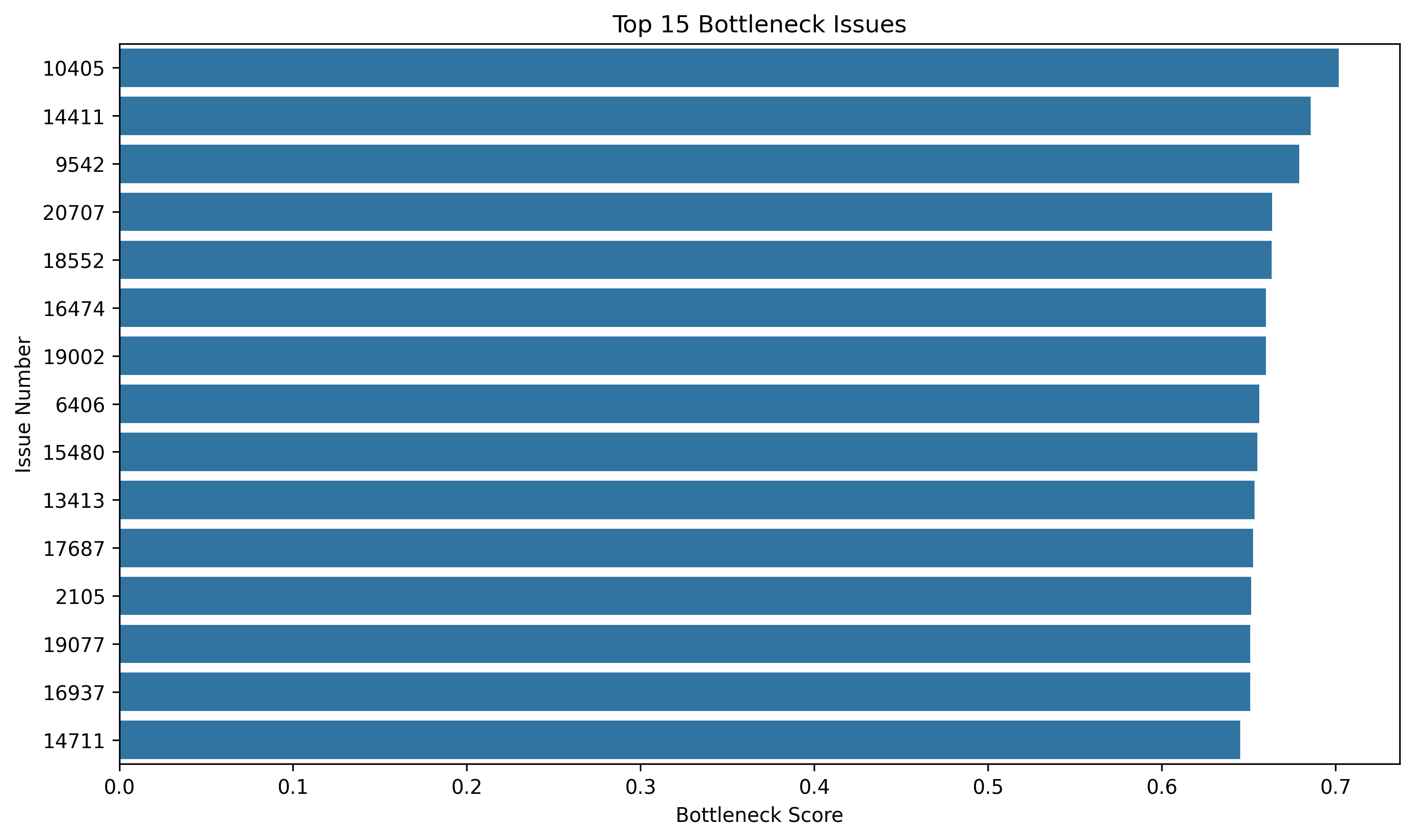}
    \caption{Top-ranked issues by bottleneck score across open and closed threads.}
    \label{fig:top_issues}
\end{figure}

Figure~\ref{fig:top_issues} ranks the highest-scoring issues from both open and closed threads. The top issue reaches a bottleneck score of approximately 0.70, with the remaining high-priority issues ranging from 0.65 to 0.67. These represent discussions where multiple warning signals co-occur and where targeted maintainer review would likely yield the highest return.

\section{Conclusion}

This paper presented SentTrack, a dual-lens framework for detecting socio-technical bottlenecks in GitHub repositories through automated analysis of issue thread conversations. Applied to the AvaloniaUI open-source repository across approximately 9,000 issue threads, the system demonstrated that real-time conversational data can surface workflow friction that traditional labels and periodic reviews consistently miss.

The horizontal pipeline addressed RQ3 by introducing an intent-focused LLM summarization stage that isolates human narrative from machine-generated noise in mixed-media issue text, enabling downstream topic extraction to reflect actual developer concerns rather than code artifacts. The pipeline produced 613 semantic clusters from the first 3,608 issues. However, the discovery-rate analysis exposed a structural limitation: the LLM's creative phrasing generates roughly two unique topic variants per issue, resulting in far more clusters than are actionable for a repository of AvaloniaUI's scale. Constraining the extraction vocabulary or introducing a dedicated topic-reduction step after clustering are the most direct paths to resolving this fragmentation.

The vertical pipeline addressed RQ1 and RQ2 by demonstrating that comment-level ABCDE classification produces interpretable thread-level signals without requiring a large language model. Across the full corpus, 49\% of threads ended in stagnation and only 13\% reached resolution, with questions frequently unanswered and concerns rarely propagating into constructive follow-up. These patterns affect downstream pipeline quality as a cross-stage effect, confirming that conversation structure is as important as text content for NLP reliability. The bottleneck detection engine showed that the dominant risk pattern is not emotional negativity but rather issues that remain open while accumulating long, unproductive discussion, with Stagnant and Long and Unresolved as the most frequently detected compound type.

Together, these results support the core premise of SentTrack: combining horizontal topic modeling with vertical thread analysis produces earlier and more interpretable signals of workflow breakdown than code metrics or manual labels alone. The primary path forward is consolidating the cluster space through a topic-reduction phase, after which bottleneck scores can be mapped directly to specific technical themes and give maintainers a unified view of both what is breaking down and where in the community it is happening.

\bibliographystyle{IEEEtran}
\bibliography{refrence}

@article{pang2008opinion,
  author  = {Pang, Bo and Lee, Lillian},
  title   = {Opinion Mining and Sentiment Analysis},
  journal = {Foundations and Trends in Information Retrieval},
  volume  = {2},
  number  = {1--2},
  pages   = {1--135},
  year    = {2008},
  doi     = {10.1561/1500000011}
}

@book{liu2012sentiment,
  author    = {Liu, Bing},
  title     = {Sentiment Analysis and Opinion Mining},
  publisher = {Morgan \& Claypool Publishers},
  year      = {2012},
  series    = {Synthesis Lectures on Human Language Technologies},
  doi       = {10.2200/S00416ED1V01Y201204HLT016}
}

@article{jurado2015sentiment,
  author  = {Jurado, Francisco and Rodriguez, Pilar},
  title   = {Sentiment Analysis in Monitoring Software Development Processes: An Exploratory Case Study on {GitHub}'s Project Issues},
  journal = {Journal of Systems and Software},
  volume  = {104},
  pages   = {82--89},
  year    = {2015},
  doi     = {10.1016/j.jss.2015.02.055}
}

@inproceedings{guzman2014commit,
  author    = {Guzman, Emitza and Az{\'o}car, David and Li, Yang},
  title     = {Sentiment Analysis of Commit Comments in {GitHub}: An Empirical Study},
  booktitle = {Proceedings of the 11th Working Conference on Mining Software Repositories},
  pages     = {352--355},
  year      = {2014},
  publisher = {ACM},
  doi       = {10.1145/2597073.2597118}
}

@inproceedings{pletea2014security,
  author    = {Pletea, Daniel and Vasilescu, Bogdan and Serebrenik, Alexander},
  title     = {Security and Emotion: Sentiment Analysis of Security Discussions on {GitHub}},
  booktitle = {Proceedings of the 11th Working Conference on Mining Software Repositories},
  pages     = {348--351},
  year      = {2014},
  publisher = {ACM},
  doi       = {10.1145/2597073.2597117}
}

@inproceedings{ortu2015bullies,
  author    = {Ortu, Marco and Adams, Bram and Destefanis, Giuseppe and Tourani, Parastou and Marchesi, Michele and Tonelli, Roberto},
  title     = {Are Bullies More Productive? Empirical Study of Affectiveness vs. Issue Fixing Time},
  booktitle = {2015 IEEE/ACM 12th Working Conference on Mining Software Repositories},
  pages     = {303--313},
  year      = {2015},
  publisher = {IEEE},
  doi       = {10.1109/MSR.2015.35}
}

@inproceedings{ortu2016emotional,
  author    = {Ortu, Marco and Murgia, Alessandro and Destefanis, Giuseppe and Tourani, Parastou and Tonelli, Roberto and Marchesi, Michele and Adams, Bram},
  title     = {The Emotional Side of Software Developers in {JIRA}},
  booktitle = {Proceedings of the 13th International Conference on Mining Software Repositories},
  pages     = {480--483},
  year      = {2016},
  publisher = {ACM},
  doi       = {10.1145/2901739.2903505}
}

@article{islam2018sentistrengthse,
  author  = {Islam, Md Rakibul and Zibran, Minhaz F.},
  title   = {{SentiStrength-SE}: Exploiting Domain Specificity for Improved Sentiment Analysis in Software Engineering Text},
  journal = {Journal of Systems and Software},
  volume  = {145},
  pages   = {125--146},
  year    = {2018},
  doi     = {10.1016/j.jss.2018.08.030}
}

@article{calefato2018senti4sd,
  author  = {Calefato, Fabio and Lanubile, Filippo and Maiorano, Federico and Novielli, Nicole},
  title   = {Sentiment Polarity Detection for Software Development},
  journal = {Empirical Software Engineering},
  volume  = {23},
  number  = {3},
  pages   = {1352--1382},
  year    = {2018},
  doi     = {10.1007/s10664-017-9546-9}
}

@inproceedings{ahmed2017senticr,
  author    = {Ahmed, Toufique and Bosu, Amiangshu and Iqbal, Anindya and Rahimi, Shahram},
  title     = {{SentiCR}: A Customized Sentiment Analysis Tool for Code Review Interactions},
  booktitle = {2017 32nd IEEE/ACM International Conference on Automated Software Engineering},
  pages     = {106--111},
  year      = {2017},
  publisher = {IEEE},
  doi       = {10.1109/ASE.2017.8115623}
}

@article{jongeling2017negative,
  author  = {Jongeling, Robbert and Sarkar, Proshanta and Datta, Subhajit and Serebrenik, Alexander},
  title   = {On Negative Results When Using Sentiment Analysis Tools for Software Engineering Research},
  journal = {Empirical Software Engineering},
  volume  = {22},
  pages   = {2543--2584},
  year    = {2017},
  doi     = {10.1007/s10664-016-9493-x}
}

@article{novielli2021assessment,
  author  = {Novielli, Nicole and Calefato, Fabio and Lanubile, Filippo and Serebrenik, Alexander},
  title   = {Assessment of Off-the-Shelf {SE}-Specific Sentiment Analysis Tools: An Extended Replication Study},
  journal = {Empirical Software Engineering},
  volume  = {26},
  number  = {4},
  pages   = {77},
  year    = {2021},
  doi     = {10.1007/s10664-021-09960-w}
}

@inproceedings{hutto2014vader,
  author    = {Hutto, C. J. and Gilbert, Eric},
  title     = {{VADER}: A Parsimonious Rule-Based Model for Sentiment Analysis of Social Media Text},
  booktitle = {Proceedings of the International AAAI Conference on Web and Social Media},
  volume    = {8},
  number    = {1},
  pages     = {216--225},
  year      = {2014},
  doi       = {10.1609/icwsm.v8i1.14550}
}

@article{colavito2024impact,
  author  = {Colavito, Giuseppe and Lanubile, Filippo and Novielli, Nicole and Quaranta, Luigi},
  title   = {Impact of Data Quality for Automatic Issue Classification Using Pre-Trained Language Models},
  journal = {Journal of Systems and Software},
  volume  = {210},
  pages   = {111838},
  year    = {2024},
  doi     = {10.1016/j.jss.2023.111838}
}

@article{hirsch2022denoising,
  author  = {Hirsch, Thomas and Hofer, Birgit},
  title   = {Detecting Non-Natural Language Artifacts for De-Noising Bug Reports},
  journal = {Automated Software Engineering},
  volume  = {29},
  number  = {2},
  pages   = {52},
  year    = {2022},
  doi     = {10.1007/s10515-022-00350-0}
}

@article{barua2014developers,
  author  = {Barua, Anton and Thomas, Stephen W. and Hassan, Ahmed E.},
  title   = {What Are Developers Talking About? An Analysis of Topics and Trends in {Stack Overflow}},
  journal = {Empirical Software Engineering},
  volume  = {19},
  number  = {3},
  pages   = {619--654},
  year    = {2014},
  doi     = {10.1007/s10664-012-9231-y}
}

@article{grootendorst2022bertopic,
  author  = {Grootendorst, Maarten},
  title   = {{BERTopic}: Neural Topic Modeling with a Class-Based {TF-IDF} Procedure},
  journal = {arXiv preprint arXiv:2203.05794},
  year    = {2022},
  doi     = {10.48550/arXiv.2203.05794}
}

@article{mcinnes2018umap,
  author  = {McInnes, Leland and Healy, John and Melville, James},
  title   = {{UMAP}: Uniform Manifold Approximation and Projection for Dimension Reduction},
  journal = {arXiv preprint arXiv:1802.03426},
  year    = {2018},
  doi     = {10.48550/arXiv.1802.03426}
}

@inproceedings{campello2013density,
  author    = {Campello, Ricardo J. G. B. and Moulavi, Davoud and Sander, J{\"o}rg},
  title     = {Density-Based Clustering Based on Hierarchical Density Estimates},
  booktitle = {Advances in Knowledge Discovery and Data Mining},
  pages     = {160--172},
  year      = {2013},
  publisher = {Springer},
  doi       = {10.1007/978-3-642-37456-2_14}
}

@article{stavrianou2009discussion,
  author  = {Stavrianou, Anna and Velcin, Julien and Chauchat, Jean-Hugues},
  title   = {A Combination of Opinion Mining and Social Network Techniques for Discussion Analysis},
  journal = {Revue des Nouvelles Technologies de l'Information},
  volume  = {RNTI-E-17},
  pages   = {25--44},
  year    = {2009}
}

@article{ravi2025threading,
  author  = {Ravi, Prerna and Lee, Dong Won and Flamia, Beatriz and David, Jasmine and Hanks, Brandon and Breazeal, Cynthia and Anderson, Emma and Lin, Grace},
  title   = {Leveraging Large Language Models to Identify Conversation Threads in Collaborative Learning},
  journal = {arXiv preprint arXiv:2510.22844},
  year    = {2025},
  doi     = {10.48550/arXiv.2510.22844}
}

@inproceedings{fan2023llmse,
  author    = {Fan, Angela and Gokkaya, Beliz and Harman, Mark and Lyubarskiy, Mitya and Sengupta, Shubho and Yoo, Shin and Zhang, Jie M.},
  title     = {Large Language Models for Software Engineering: Survey and Open Problems},
  booktitle = {2023 IEEE/ACM International Conference on Software Engineering: Future of Software Engineering},
  pages     = {31--53},
  year      = {2023},
  publisher = {IEEE},
  doi       = {10.1109/ICSE-FoSE59343.2023.00008}
}

@inproceedings{lewis2020rag,
  author    = {Lewis, Patrick and Perez, Ethan and Piktus, Aleksandra and Petroni, Fabio and Karpukhin, Vladimir and Goyal, Naman and K{\"u}ttler, Heinrich and Lewis, Mike and Yih, Wen-tau and Rockt{\"a}schel, Tim and Riedel, Sebastian and Kiela, Douwe},
  title     = {Retrieval-Augmented Generation for Knowledge-Intensive {NLP} Tasks},
  booktitle = {Advances in Neural Information Processing Systems},
  volume    = {33},
  pages     = {9459--9474},
  year      = {2020}
}

@inproceedings{correia2024devmentorai,
  author    = {Correia, Jo{\~a}o and Nicholson, Morgan C. and Coutinho, Daniel and Barbosa, Caio and Castelluccio, Marco and Gerosa, Marco and Garcia, Alessandro and Steinmacher, Igor},
  title     = {Unveiling the Potential of a Conversational Agent in Developer Support: Insights from Mozilla's {PDF.js} Project},
  booktitle = {Proceedings of the 1st ACM International Conference on AI-Powered Software},
  pages     = {10--18},
  year      = {2024},
  publisher = {ACM},
  doi       = {10.1145/3664646.3664758}
}

@article{correia2025firefox,
  author  = {Correia, Jo{\~a}o and Coutinho, Daniel and Castelluccio, Marco and Barbosa, Caio and de Mello, Rafael and Sarma, Anita and Garcia, Alessandro and Gerosa, Marco and Steinmacher, Igor},
  title   = {A Comparison of Conversational Models and Humans in Answering Technical Questions: The {Firefox} Case},
  journal = {arXiv preprint arXiv:2510.21933},
  year    = {2025},
  doi     = {10.48550/arXiv.2510.21933}
}

@article{imran2022sentimenttools,
  author  = {Imran, Muhammad and Menon, Anjila and Abubakar, Muhammad and Mahmood, Wasif and Niazi, Mahmood},
  title   = {Sentiment Analysis Tools in Software Engineering: A Systematic Mapping Study},
  journal = {Information and Software Technology},
  volume  = {151},
  pages   = {107018},
  year    = {2022},
  doi     = {10.1016/j.infsof.2022.107018}
}

@inproceedings{coutinho2024looksgood,
  author    = {Coutinho, Daniel and Cito, Luisa and Lima, Maria Vit{\'o}ria and Arantes, Beatriz and Alves Pereira, Juliana and Arriel, Johny and Godinho, Jo{\~a}o and Martins, Vinicius and Lib{\'o}rio, Paulo V{\'i}tor C. F. and Leite, Leonardo and Garcia, Alessandro and Assun{\c c}{\~a}o, Wesley K. G. and Steinmacher, Igor and Baffa, Augusto and Fonseca, Baldoino},
  title     = {{``}Looks Good To Me ;-){''}: Assessing Sentiment Analysis Tools for Pull Request Discussions},
  booktitle = {Proceedings of the 28th International Conference on Evaluation and Assessment in Software Engineering},
  pages     = {211--221},
  year      = {2024},
  publisher = {ACM},
  doi       = {10.1145/3661167.3661189}
}

@inproceedings{finkel2006cascading,
  author    = {Finkel, Jenny Rose and Manning, Christopher D. and Ng, Andrew Y.},
  title     = {Solving the Problem of Cascading Errors: Approximate Bayesian Inference for Linguistic Annotation Pipelines},
  booktitle = {Proceedings of the 2006 Conference on Empirical Methods in Natural Language Processing},
  pages     = {618--626},
  year      = {2006},
  publisher = {Association for Computational Linguistics}
}

@inproceedings{caselli2015error,
  author    = {Caselli, Tommaso and Vossen, Piek and van Erp, Marieke and Fokkens, Antske and Ilievski, Filip and Izquierdo, Rub{\'e}n and Le, Minh and Morante, Roser and Postma, Marten},
  title     = {When It's All Piling Up: Investigating Error Propagation in an {NLP} Pipeline},
  booktitle = {Proceedings of the 20th International Conference on Applications of Natural Language to Information Systems},
  pages     = {417--428},
  year      = {2015},
  publisher = {Springer},
  doi       = {10.1007/978-3-319-19581-0_34}
}

@inproceedings{lin2004rouge,
  author    = {Lin, Chin-Yew},
  title     = {{ROUGE}: A Package for Automatic Evaluation of Summaries},
  booktitle = {Text Summarization Branches Out: Proceedings of the ACL-04 Workshop},
  pages     = {74--81},
  year      = {2004},
  publisher = {Association for Computational Linguistics}
}

@inproceedings{kryscinski2020factcc,
  author    = {Kry{\'s}ci{\'n}ski, Wojciech and McCann, Bryan and Xiong, Caiming and Socher, Richard},
  title     = {Evaluating the Factual Consistency of Abstractive Text Summarization},
  booktitle = {Proceedings of the 2020 Conference on Empirical Methods in Natural Language Processing},
  pages     = {9332--9346},
  year      = {2020},
  publisher = {Association for Computational Linguistics},
  doi       = {10.18653/v1/2020.emnlp-main.750}
}

@inproceedings{clark2023seahorse,
  author    = {Clark, Elizabeth and Rijhwani, Shruti and Gehrmann, Sebastian and Maynez, Joshua and Aharoni, Roee and Nikolaev, Vitaly and Sellam, Thibault and Siddhant, Aditya and Das, Dipanjan and Parikh, Ankur},
  title     = {{SEAHORSE}: A Multilingual, Multifaceted Dataset for Summarization Evaluation},
  booktitle = {Proceedings of the 2023 Conference on Empirical Methods in Natural Language Processing},
  pages     = {9397--9413},
  year      = {2023},
  publisher = {Association for Computational Linguistics},
  doi       = {10.18653/v1/2023.emnlp-main.584}
}

@article{fu2023gptscore,
  author  = {Fu, Jinlan and Ng, See-Kiong and Jiang, Zhengbao and Liu, Pengfei},
  title   = {{GPTScore}: Evaluate as You Desire},
  journal = {arXiv preprint arXiv:2302.04166},
  year    = {2023},
  doi     = {10.48550/arXiv.2302.04166}
}

@inproceedings{augenstein2017scienceie,
  author    = {Augenstein, Isabelle and Das, Mrinal and Riedel, Sebastian and Vikraman, Lakshmi and McCallum, Andrew},
  title     = {{SemEval} 2017 Task 10: {ScienceIE} -- Extracting Keyphrases and Relations from Scientific Publications},
  booktitle = {Proceedings of the 11th International Workshop on Semantic Evaluation},
  pages     = {546--555},
  year      = {2017},
  publisher = {Association for Computational Linguistics},
  doi       = {10.18653/v1/S17-2091}
}

@inproceedings{moulavi2014dbcv,
  author    = {Moulavi, Davoud and Jaskowiak, Pablo A. and Campello, Ricardo J. G. B. and Zimek, Arthur and Sander, J{\"o}rg},
  title     = {Density-Based Clustering Validation},
  booktitle = {Proceedings of the 2014 SIAM International Conference on Data Mining},
  pages     = {839--847},
  year      = {2014},
  publisher = {SIAM},
  doi       = {10.1137/1.9781611973440.96}
}

@manual{b2,
  author       = {{GitHub}},
  title        = {{About Issues}},
  organization = {{GitHub Docs}},
  year         = {n.d.},
  url          = {https://docs.github.com/en/issues/tracking-your-work-with-issues/learning-about-issues/about-issues},
  note         = {Accessed: Mar. 6, 2026}
}

@manual{b3,
  author       = {{GitHub}},
  title        = {{About Discussions}},
  organization = {{GitHub Docs}},
  year         = {n.d.},
  url          = {https://docs.github.com/en/discussions/collaborating-with-your-community-using-discussions/about-discussions},
  note         = {Accessed: Mar. 6, 2026}
}

\end{document}